\RequirePackage{fix-cm} 
\documentclass[a4paper, twoside, reqno, 12pt, dvips]{amsart}
\usepackage{fixltx2e}     

\usepackage[english]{babel}
\usepackage[latin1]{inputenc}

\usepackage{eucal}
\usepackage{esint}
\usepackage{amsgen}
\usepackage{amsthm}
\usepackage{amssymb}
\usepackage{amsmath}
\usepackage{amsfonts}
\usepackage{verbatim}
\usepackage[nice]{nicefrac}
\usepackage{mathrsfs, dsfont}

\usepackage{a4wide}
\usepackage{xspace}

\usepackage{indentfirst}
\usepackage{graphicx}
\usepackage{subfigure}
\usepackage[section]{placeins}
\usepackage{psfrag}

\usepackage{microtype}
\usepackage[bf, up, hang]{caption}

\usepackage[usenames, dvipsnames]{pstricks}
\usepackage{epsfig}
\usepackage{pst-grad} 
\usepackage{pst-plot} 

\usepackage{color}
\definecolor{oneblue}{rgb}{0.0, 0.0, 0.85}
\definecolor{darkgrey}{rgb}{0.273, 0.281, 0.30}

\usepackage[colorlinks,
            urlcolor=oneblue,
            linkcolor=oneblue,
            citecolor=oneblue,
            bookmarksopen=false,
            pagebackref]{hyperref}

\usepackage[compact]{titlesec}

\titleformat{\section}{\normalfont\Large\bfseries\sffamily\center\color{darkgrey}}{\thesection.}{0.5em}{}{}
\titleformat{\subsection}{\normalfont\large\bfseries\sffamily\color{darkgrey}}{\thesubsection.}{0.4em}{}{}
\titleformat{\subsubsection}{\normalfont\normalsize\bfseries\sffamily\color{darkgrey}}{\thesubsubsection.}{0.3em}{}{}

\titlespacing*{\section}{1.0em}{1.0em}{0.8em}[0em]
\titlespacing*{\subsection}{1.0em}{1.0em}{0.8em}[0em]
\titlespacing*{\subsubsection}{1.0em}{0.7em}{0.6em}[0em]

\usepackage{titletoc}

\contentsmargin{0.0em}
\dottedcontents{section}[2.5em]{\addvspace{0.45em}\bfseries}{1.0em}{0pc}
\dottedcontents{subsection}[4.5em]{}{2.6em}{0pc}
\dottedcontents{subsubsection}[5.5em]{}{3.0em}{0pc}

\usepackage{fancyhdr}
\usepackage{lastpage}

\newcommand*\Title{Derivation of dissipative Boussinesq equations using the D2N operator}
\newcommand*\Authors{D.~Dutykh \& O.~Goubet}

\usepackage{acronym}

\numberwithin{equation}{section}

\newtheorem{remark}{Remark}

\newcommand{\n}{\hat{n}}
\newcommand{\ub}{\bar{u}}
\newcommand{\x}{\mathbf{x}}
\newcommand{\phis}{\varphi}
\newcommand{\R}{\mathbb{R}}
\newcommand{\D}{\mathcal{D}}
\newcommand{\phih}{\hat{\phi}}
\newcommand{\phib}{\bar{\phi}}
\newcommand{\psit}{\tilde{\psi}}
\newcommand{\Psit}{\tilde{\Psi}}
\newcommand{\phit}{\tilde{\phi}}
\newcommand{\Phit}{\tilde{\Phi}}
\newcommand{\phish}{\hat{\varphi}}
\newcommand{\phibh}{\hat{\bar{\phi}}}
\newcommand{\dt}{\boldsymbol{\partial}_t}
\newcommand{\dz}{\boldsymbol{\partial}_z}
\renewcommand{\u}{\mathbf{u}}
\newcommand{\eps}{\varepsilon}
\renewcommand{\O}{\mathcal{O}}

\renewcommand{\div}{\grad\scal}
\newcommand{\Dd}{\tilde{\mathcal{D}}}
\newcommand{\scal}{\boldsymbol{\cdot}}
\newcommand{\grad}{\boldsymbol{\nabla}}
\newcommand{\half}{{\textstyle{1\over2}}}

\newcommand{\pd}[2]{\frac{\partial#1}{\partial#2}}

\begin{document}

\title[\Title]{Derivation of dissipative Boussinesq equations using the Dirichlet-to-Neumann operator approach}

\author[D.~Dutykh]{Denys Dutykh$^*$}
\address{University College Dublin, School of Mathematical Sciences, Belfield, Dublin 4, Ireland \and LAMA, UMR 5127 CNRS, Universit\'e de Savoie, Campus Scientifique, 73376 Le Bourget-du-Lac Cedex, France}
\email{Denys.Dutykh@ucd.ie}
\urladdr{http://www.denys-dutykh.com/}
\thanks{$^*$ Corresponding author}

\author[O. Goubet]{Olivier Goubet}
\address{LAMFA CNRS UMR 7352, Universit\'e de Picardie Jules Verne,
33 rue Saint-Leu 80039 Amiens cedex}
\email{Olivier.Goubet@u-picardie.fr}
\urladdr{http://www.lamfa.u-picardie.fr/goubet/}

\begin{abstract}
The water wave theory traditionally assumes the fluid to be perfect, thus neglecting all effects of the viscosity. However, the explanation of several experimental data sets requires the explicit inclusion of dissipative effects. In order to meet these practical problems, the theory of visco-potential flows has been developed (see P.-F.~\textsc{Liu} \& A.~\textsc{Orfila} \cite{Liu2004} and D.~\textsc{Dutykh} \& F.~\textsc{Dias} \cite{DutykhDias2007}). Then, usually this formulation is further simplified by developing the potential in an entire series in the vertical coordinate and by introducing thus, the long wave approximation. In the present study we propose a derivation of dissipative Boussinesq equations which is based on two various asymptotic expansions of the Dirichlet-to-Neumann (D2N) operator. Both employed methods yield the same system at the leading order by different ways.

\bigskip
\noindent \textbf{\keywordsname:} Boussinesq equations; viscosity; dissipation; dispersive waves; boundary layer
\end{abstract}


\maketitle
\tableofcontents
\thispagestyle{empty}

\section{Introduction}

The water wave problem has a long history which counts more than two hundred years \cite{Craik2004}. Traditionally the fluid under consideration is supposed to be perfect which allowed to reveal several useful variational structures \cite{Petrov1964, Luke1967, Zakharov1968, Clamond2009}. However this assumption is not valid for all time scales and for all physical situations. The explanation of several laboratory experiments and some modeling problems require the explicit inclusion of weak dissipative effects. In the prominent work by J.~Bona \emph{et al.} (1981) \cite{Bona1981} one can find the following passage emphasizing the importance of dissipative effects:
\begin{quote}
  ``\dots it was found that the inclusion of a dissipative term was much more important than the inclusion of the nonlinear term, although the inclusion of the nonlinear term was undoubtedly beneficial in describing the observations.''
\end{quote}
Another more recent experimental study emphasizes more specifically the boundary layer effects \cite{Hammack2004}:
\begin{quote}
``A third source of experimental differences is \emph{residual boundary layer} motions that are left behind as a solitary wave propagates in the channel. In both the co- and counterpropagating binary collision experiments solitary waves encounter the boundary-layer wakes of the other wave. This wake does have a small effect on wave speeds that can be significant in our data analysis. Detailed measurements of these boundary layer motions were not made; hence, it is not known how reproducible they are.''
\end{quote}

The complete model of viscous free-surface flows should consider two phases (air and water) separated by an interface. Each phase is described by incompressible Navier-Stokes equations with respective viscosities. Such numerical models currently exist \cite{Raval2009, Fuster2009}, however they allow to simulate only a few wavelengthes in reasonable times. Consequently, several approximate models have been developed to take into account weak viscous effects. One of the first simplifications consists in replacing the upper light fluid by a free surface. This research direction was successfully pursued in several studies mentioned below.

Historically various dissipative terms have been incorporated in long wave equations such as Burgers \cite{Sugimoto1991}, KdV \cite{Ott1970, KM, Grimshaw2003} or some Boussinesq type systems \cite{Khabakhpashev1987, Khabakhpashev1997, CG, Dutykh2007, Dutykh2007a, Khabakhpashev2008, Dutykh2010f, Chen2010}. Some attempt has been also made to incorporate ad-hoc dissipative terms into the complete water wave problem as well \cite{Lundgren1989, Ruvinsky1991, Skandrani, Joseph2006, Wu2006, Kharif2010}.

Recently, the theory of visco-potential flows has been developed for weakly dissipative regimes \cite{Liu2004, DutykhDias2007, Liu2007, Dutykh2008a, Dutykh2008b}. The main idea consists in writing the Helmholtz decomposition of the velocity field for linearized Navier--Stokes equations with free surface. These equations can be identically satisfied by a potential velocity field but not the boundary conditions. Then, the rotational part is expressed asymptotically (and hence, approximatively) in terms of the velocity potential and free surface elevation. In this way we introduce some corrections to boundary conditions on the free surface and/or on the bottom. In deep water case this question was studied by F.~\textsc{Dias} \emph{et al.} (2008) \cite{Dias2007}. In the finite depth case the situation is more complex since a boundary layer at the bottom should be taken into account \cite{Liu2004, DutykhDias2007}. This physical effect modifies the bottom boundary condition by adding a nonlocal term in time which represents mathematically a half-order integral. Consequently, a memory effect is introduced into the model in the finite depth. Some attempt to the efficient discretization of this fractional integral has been performed during last years \cite{Torsvik2007, Dutykh2008a, Chen2010, Chehab2011}.

The main scope of the present study is to combine the visco-potential approach with Dirichlet-to-Neumann (D2N) operator formulation of the water wave problem \cite{Petrov1964, Zakharov1968, Craig1992, Craig1993} which can be successfully used for water wave modeling purposes \cite{Craig1994, Craig2005, Lannes2009}. Using this novel visco-potential formulation and an asymptotic expansion of the D2N operator, we derive a long wave model of Boussinesq type. We underline that our derivation remains at the formal level. More theoretical study will be needed to justify our current developments.

The present study is organized as follows. In Section~\ref{sec:models} we present the visco-potential formulation with D2N operator, while dissipative Boussinesq equations are derived in Section~\ref{sec:dissB} following two different approaches presented in Sections~\ref{sec:shape} and \ref{sec:lwave}. Finally, this study is ended by outlining main conclusions in Section \ref{sec:concl}. In Section~\ref{sec:4} we discuss some results related to dissipative Boussinesq system and the dispersion relation analysis.

\section{Mathematical modeling}\label{sec:models}

\begin{figure}
\begin{center}
\scalebox{1} 
{
\begin{pspicture}(0,-3.6097188)(15.559063,3.6367188)
\definecolor{color31b}{rgb}{0.2901960784313726,0.19607843137254902,0.19607843137254902}
\definecolor{color31}{rgb}{0.2,0.043137254901960784,0.043137254901960784}
\definecolor{color142}{rgb}{0.050980392156862744,0.13725490196078433,0.7490196078431373}
\psline[linewidth=0.06,linecolor=color31,fillstyle=solid,fillcolor=color31b](0.04,-2.8367188)(3.2,-2.8167188)(5.62,-2.0167189)(8.82,-2.0367188)(11.22,-2.7767189)(15.4,-2.7567186)
\psline[linewidth=0.025999999cm,arrowsize=0.05291667cm 2.0,arrowlength=1.4,arrowinset=0.4]{->}(0.04,1.5832813)(15.26,1.6032813)
\psline[linewidth=0.025999999cm,arrowsize=0.05291667cm 2.0,arrowlength=1.4,arrowinset=0.4]{<-}(7.24,3.5832813)(7.24,-3.5967188)
\usefont{T1}{ptm}{m}{n}
\rput(15.124531,1.2332813){$\x$}
\usefont{T1}{ptm}{m}{n}
\rput(6.9845314,3.4332812){$z$}
\usefont{T1}{ptm}{m}{n}
\rput(6.934531,1.3732812){$O$}
\psline[linewidth=0.025999999cm,arrowsize=0.05291667cm 2.0,arrowlength=1.4,arrowinset=0.4]{<->}(10.42,-2.4967186)(10.42,1.5832813)
\usefont{T1}{ptm}{m}{n}
\rput(9.684531,-0.36671874){$h(\x,t)$}
\pscustom[linewidth=0.06,linecolor=color142]
{
\newpath
\moveto(0.0,1.1632812)
\lineto(0.21,1.2432812)
\curveto(0.315,1.2832812)(0.525,1.3832812)(0.63,1.4432813)
\curveto(0.735,1.5032812)(0.95,1.5932813)(1.06,1.6232812)
\curveto(1.17,1.6532812)(1.44,1.7182813)(1.6,1.7532812)
\curveto(1.76,1.7882812)(2.055,1.8232813)(2.19,1.8232813)
\curveto(2.325,1.8232813)(2.62,1.7982812)(2.78,1.7732812)
\curveto(2.94,1.7482812)(3.265,1.6732812)(3.43,1.6232812)
\curveto(3.595,1.5732813)(3.925,1.5182812)(4.09,1.5132812)
\curveto(4.255,1.5082812)(4.615,1.5232812)(4.81,1.5432812)
\curveto(5.005,1.5632813)(5.345,1.6182812)(5.49,1.6532812)
\curveto(5.635,1.6882813)(5.91,1.7332813)(6.04,1.7432812)
\curveto(6.17,1.7532812)(6.44,1.7782812)(6.58,1.7932812)
\curveto(6.72,1.8082813)(7.005,1.8332813)(7.15,1.8432813)
\curveto(7.295,1.8532813)(7.62,1.8532813)(7.8,1.8432813)
\curveto(7.98,1.8332813)(8.305,1.7982812)(8.45,1.7732812)
\curveto(8.595,1.7482812)(8.905,1.6932813)(9.07,1.6632812)
\curveto(9.235,1.6332812)(9.625,1.6232812)(9.85,1.6432812)
\curveto(10.075,1.6632812)(10.435,1.7182813)(10.57,1.7532812)
\curveto(10.705,1.7882812)(10.995,1.8582813)(11.15,1.8932812)
\curveto(11.305,1.9282813)(11.625,1.9882812)(11.79,2.0132813)
\curveto(11.955,2.0382812)(12.385,2.0682812)(12.65,2.0732813)
\curveto(12.915,2.0782812)(13.31,2.0632813)(13.44,2.0432813)
\curveto(13.57,2.0232813)(13.82,1.9782813)(13.94,1.9532813)
\curveto(14.06,1.9282813)(14.305,1.8932812)(14.43,1.8832812)
\curveto(14.555,1.8732812)(14.87,1.8632812)(15.06,1.8632812)
}
\psline[linewidth=0.04cm,arrowsize=0.05291667cm 2.0,arrowlength=1.4,arrowinset=0.4]{<->}(2.24,1.8032813)(2.24,-2.7767189)
\usefont{T1}{ptm}{m}{n}
\rput(1.4545312,-0.50671875){$H(\x,t)$}
\psline[linewidth=0.025999999cm,arrowsize=0.05291667cm 2.0,arrowlength=1.4,arrowinset=0.4]{<->}(12.82,2.0832813)(12.82,1.6032813)
\psline[linewidth=0.025999999cm](12.82,2.0632813)(12.82,2.6432812)
\usefont{T1}{ptm}{m}{n}
\rput(12.134531,2.3932812){$\eta(\x,t)$}
\end{pspicture}
}
\caption{\em\small Sketch of the physical domain.}
\label{fig:sketch}
\end{center}
\end{figure}
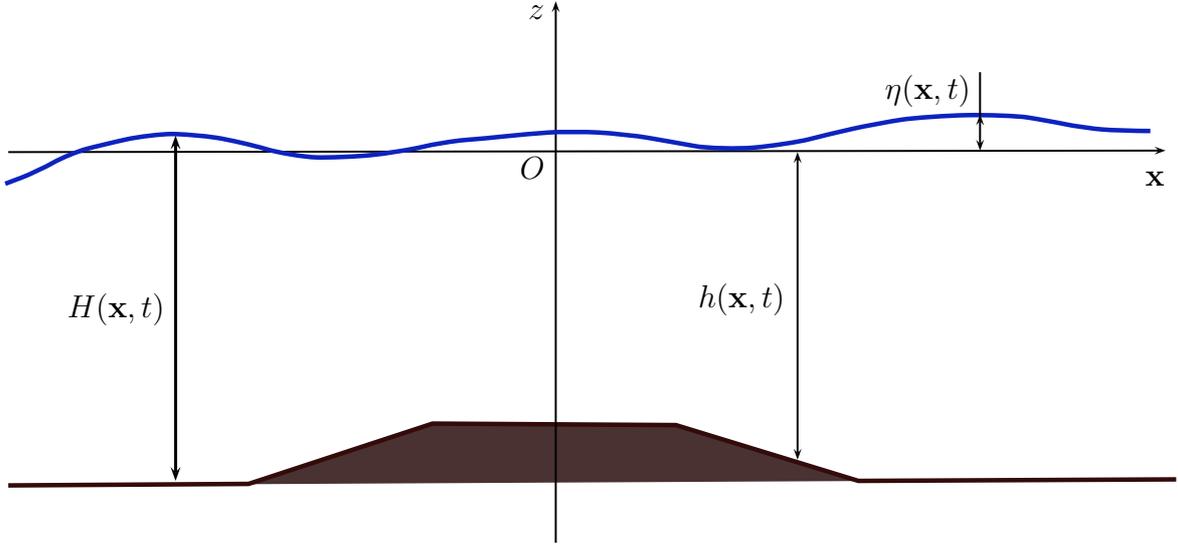

Consider an ideal incompressible fluid of constant density $\rho$. The vertical projection of the fluid domain $\Omega$ is a subset of $\R^2$. The horizontal independent variables are denoted by $\x = (x,y)$ and the vertical one by $z$. The origin of the cartesian coordinate system is chosen such that the surface $z=0$ corresponds to the still water level. The fluid is bounded below by the bottom $z = -h(\x,t)$ and above by the free surface $z = \eta (\x,t)$. We assume that the total depth $H(\x, t) := h(\x,t) + \eta (\x,t)$ remains positive $H (\x,t) \geq h_0 > 0$ at all times $t \in [0, T]$. The sketch of the physical domain is shown in Figure~\ref{fig:sketch}.

\begin{remark}
We make the classical assumption that the free surface is a graph $z = \eta (\x,t)$ of a single-valued function. It means in practice that we exclude some interesting phenomena, (e.g. wave breaking phenomena) which are out of the scope of this modeling paradigm.
\end{remark}

The governing equations of the classical water wave problem are the following \cite{Lamb1932, Stoker1958, Mei1994, Whitham1999}:
\begin{eqnarray}
  \grad^2\phi + \boldsymbol{\partial}^2_{zz}\phi &=& 0,
  \qquad (\x, z) \in \Omega\times [-h, \eta], \label{eq:laplace} \\
  \dt\eta + \grad\phi\scal\grad\eta - \dz\phi &=& 0,
  \qquad z = \eta(\x, t), \label{eq:kinematic} \\
  \dt\phi + \half|\grad\phi|^2 + \half(\dz\phi)^2 + g\eta &=& 0,
  \qquad z = \eta(\x,t), \label{eq:bernoulli} \\
  \dt h + \grad\phi\scal\grad h + \dz\phi &=& 0,
  \qquad z = -h(\x,t), \label{eq:bottomkin}
\end{eqnarray}
with $\phi$ the velocity potential (by definition the velocity field $(\u, v) = \grad_{x,z}\phi$), $g$ the acceleration due to gravity force and $\grad = (\boldsymbol{\partial}_x, \boldsymbol{\partial}_y)$ denotes the gradient operator in horizontal Cartesian coordinates.

Incompressibility leads to Laplace's equation for $\phi$. The main difficulty of the water wave problem lies on the nonlinear free boundary conditions. Equations \eqref{eq:kinematic} and \eqref{eq:bottomkin} express the free-surface kinematic condition and bottom impermeability respectively, while the dynamic condition \eqref{eq:bernoulli} expresses the free surface isobarity.

\begin{remark}
Surface tension effects can also be included in the water wave problem. In this case, the Bernoulli condition \eqref{eq:bernoulli} has to be modified (see \cite{Dias1999} for a general review of this topic):
\begin{equation*}
  \dt\phi + \half|\grad\phi|^2 + \half(\dz\phi)^2 + g\eta =
  \frac{\sigma}{\rho}\div\Bigl(\frac{\grad\eta}{\sqrt{1 + |\grad\eta|^2}}\Bigr),
  \quad z=\eta(\x,t),
\end{equation*}
where $\sigma$ is the surface tension coefficient. However, this effect is not of primary importance in our study and, consequently, will be ignored.
\end{remark}

Below we will need the unitary exterior normals to the fluid domain. It is straightforward to obtain the following expressions for the normals at the free surface and bottom respectively:
\begin{equation*}
  \n_f = \frac{1}{\sqrt{1 + |\grad\eta|^2}}\left|
    \begin{array}{c}
      -\grad\eta \\
      1
    \end{array}
  \right.,
  \qquad
  \n_b = \frac{1}{\sqrt{1 + |\grad h|^2}}\left|
    \begin{array}{c}
      -\grad h \\
      -1
    \end{array}
  \right. .
\end{equation*}

Russian physicists A.~\textsc{Petrov} (1964) \cite{Petrov1964} and V.~\textsc{Zakharov} (1968) \cite{Zakharov1968} proposed a more compact formulation of the water wave problem \eqref{eq:laplace} -- \eqref{eq:bottomkin} based on the Hamiltonian structure. Let us introduce the trace of the velocity potential at the free surface:
\begin{equation*}
  \phis(\x, t) := \phi(\x, \eta(\x,t), t).
\end{equation*}
This variable plays the role of generalized momentum in the Hamiltonian description of water waves \cite{Zakharov1968, Dias2006a}. The second canonical variable is the free surface elevation $\eta$. Another important ingredient is the normal velocity at the free surface $v_n$ which is defined as:
\begin{equation}\label{eq:normalv}
  v_n (\x,t) := \sqrt{1 + |\grad\eta|^2}\left.\pd{\phi}{\n_f}\right|_{z=\eta} =
  \left.(\dz\phi - \grad\phi\scal\grad\eta)\right|_{z=\eta}.
\end{equation}

Using all these elements, the boundary conditions (\ref{eq:kinematic}) and (\ref{eq:bernoulli}) on the free surface can be rewritten in terms of $\phis$, $v_n$ and $\eta$ \cite{Craig1992, Craig1993, Fructus2005} as following:
\begin{equation}\label{eq:dynamics}
 \begin{array}{rl}
  \dt\eta - \D_\eta(\phis) &= 0, \\
  \dt\phis + \half|\grad\phis|^2 + g\eta
  - \frac{1}{2(1+|\grad\eta|^2)}\bigl[\D_\eta(\phis) +
   \grad\phis\cdot\grad\eta \bigr]^2 &= 0,
 \end{array}
\end{equation}
where we introduced the so-called classical Dirichlet-to-Neumann (D2N) $\D_\eta$ operator \cite{Coifman1985, Craig1993}, which maps the velocity potential at the free surface $\phis$ to the normal velocity $v_n$:
\begin{eqnarray*}
  \D_\eta: & \phis \mapsto v_n = \sqrt{1 + |\grad\eta|^2}
  \left.\pd{\phi}{\n_f}\right|_{z=\eta} \\
  & \left|
  \begin{array}{rl}
    \grad^2\phi + \boldsymbol{\partial}^2_{zz}\phi &= 0, \quad (\x, z) \in \Omega\times [-h, \eta], \\
    \phi &= \phis, \quad z=\eta, \\
    \sqrt{1 + |\grad h|^2}\displaystyle{\pd{\phi}{\n_b}} &= \dt h, \quad z=-h.
  \end{array}\right.
\end{eqnarray*}
This operator takes the free surface shape $\eta(\x,t)$ as a parameter and the velocity potential trace $\phis(\x,t)$ as an argument (Dirichlet boundary condition). As a result it returns the normal velocity $v_n = \sqrt{1 + |\grad\eta|^2}\left.\pd{\phi}{\n_f}\right|_{z=\eta}$ at the free surface (Neumann data) which explains the common name of this operator.

\begin{remark}
The function $h(\x,t)$ represents the ocean bathymetry (depth below the still water level, see Figure~\ref{fig:sketch}) and is assumed to be known. The dependence on time is included in order to take into account the bottom motion during, for example, an underwater earthquake \cite{Dias2006, Dutykh2006, ddk}. Since in the present study we are mainly interested in quantifying a solitary wave amplitude decay under the viscous damping, we will take hereafter the even bottom case $h(\x,t) \equiv h = const$.
\end{remark}

Now we are going to introduce weak dissipative effects directly into the Petrov-Zakharov formulation \eqref{eq:dynamics} in the spirit of the visco-potential flows theory \cite{Liu2004, DutykhDias2007, Dutykh2008b}:
\begin{equation}\label{eq:dynamicsDiss}
 \begin{array}{rl}
  \dt\eta - \Dd_\eta(\phis) &= 2\nu\grad^2\eta, \\
  \dt\phis + \half|\grad\phis|^2 + g\eta
  - \frac{1}{2(1+|\grad\eta|^2)}\bigl[\Dd_\eta(\phis) +
   \grad\phis\cdot\grad\eta \bigr]^2 &= 2\nu\grad^2\phis,
 \end{array}
\end{equation}
where $\nu$ is the kinematic (molecular or eddy) viscosity and the modified D2N operator $\Dd$ is defined as:
\begin{eqnarray*}
  \Dd_\eta: & \phis \mapsto v_n = \sqrt{1 + |\grad\eta|^2}
  \left.\pd{\phi}{\n_f}\right|_{z=\eta} \\
  & \left|
  \begin{array}{rl}
    \grad^2\phi + \boldsymbol{\partial}^2_{zz}\phi &= 0, \quad (\x, z) \in \Omega\times [-h, \eta], \\
    \phi &= \phis, \quad z=\eta, \\
    \dz\phi &=
    -\sqrt{\frac{\nu}{\pi}}\int\limits_0^t \frac{\phi_{zz}}{\sqrt{t-\tau}}\;d\tau,
    \quad z=-h.
  \end{array}\right.
\end{eqnarray*}
The local dissipative terms in \eqref{eq:dynamicsDiss} are due to viscous effects in the bulk of the fluid while the integral term in the bottom kinematic condition takes into account boundary layer effects at the bottom. For more details on the modeling and derivation of visco-potential flows formulation we refer to \cite{Liu2004, Dutykh2007a, Dutykh2008a}. In the next Section we will simplify dynamic equations \eqref{eq:dynamicsDiss} by applying a long wave approximation.

\begin{remark}\label{rem:lapl}
The bottom kinematic boundary condition can be equivalently rewritten using the Laplace equation which is still valid on boundaries under the assumption of their regularity:
\begin{equation*}
  \dz\phi = \sqrt{\frac{\nu}{\pi}}\int\limits_0^t \frac{\grad^2\phi}{\sqrt{t-s}}\;ds,
    \quad z=-h.
\end{equation*}
The notation can be greatly simplified if we apply the Laplace transform to both sides of the last identity \cite{Widder1946, gradshteyn}:
\begin{equation*}
  \dz\phi = \sqrt{\frac{\nu}{\tau}}\;\grad^2\phi,
\end{equation*}
where $\tau$ is the Laplace transform parameter (to shorten the notation the images of the Laplace transform are denoted by the same symbol).

We stress that the use of the Laplace transform is not fundamental to the subsequent derivation. Alternatively we could denote the half-integral by another symbol. We find it natural to use the apparatus of the operational calculus.
\end{remark}

\subsection{Dimensionless variables}

In order to apply an asymptotic expansion in the following Section, we have to introduce dimensionless variables to reveal small parameters \cite{BCS, Dutykh2007, DMII, Lannes2009}.

Unscaled independent variables are defined as follows:
\begin{equation*}
  \x\,' = \frac{\x}{\ell}, \qquad z' = \frac{z}{h}, \qquad
  t' = \frac{t}{\ell/\sqrt{gh}},
\end{equation*}
where $\ell$ is a typical wavelength, $h$ is the average water depth and $\sqrt{gh}$ gives the phase speed of infinitely long shallow water waves. The scaling of dependent variables needs one more characteristic length --- the typical wave amplitude $a$:
\begin{equation*}
  \eta' = \frac{\eta}{a}, \qquad
  \phi' = \frac{\phi}{\ell a g / \sqrt{gh}}.
\end{equation*}
In dimensionless variables the governing equations \eqref{eq:dynamicsDiss} take the following form (we dropped out the primes for the sake of convenience):
\begin{equation}\label{eq:kinem}
  \dt\eta - \frac{1}{\mu^2}\Dd_{\eps\eta}(\phis) = 2\nu\grad^2\eta,
\end{equation}
\begin{equation}\label{eq:dyn}
  \dt\phis + \frac{\eps}{2}|\grad\phis|^2 + \eta
  - \frac{\eps}{\mu^2}\frac{\bigl[\Dd_{\eps\eta}(\phis) +
   \eps\mu^2\grad\phis\cdot\grad\eta\bigr]^2}%
  {2(1+\eps^2\mu^2|\grad\eta|^2)}
  = 2\nu\grad^2\phis,
\end{equation}
and the D2N operator $\Dd_{\eps\eta}$ in unscaled variables becomes:
\begin{eqnarray*}
  \Dd_{\eps\eta}: \phis & \mapsto v_n = \sqrt{1 + \eps^2\mu^2|\grad\eta|^2}
  \left.\pd{\phi}{\n_f}\right|_{z=\eps\eta} =
  \left.\bigl(\dz\phi - \eps\mu^2\grad\phi\cdot\grad\eta\bigr)\right|_{z=\eps\eta} \\
  & \left|
  \begin{array}{rl}
    \grad^2\phi + \frac{1}{\mu^2}\partial^2_{zz}\phi &= 0, \quad (\x, z) \in \Omega\times [-1, \eps\eta], \\
    \phi &= \phis, \quad z=\eps\eta, \\
    \dz\phi &=
    -\sqrt{\frac{\nu}{\pi\mu^2}}\int\limits_0^t\frac{\phi_{zz}}{\sqrt{t-s}}\;ds,
    \quad z=-1.
  \end{array}\right.
\end{eqnarray*}

\begin{remark}
In the light of Remark~\ref{rem:lapl}, the bottom boundary condition can be more conveniently rewritten in dimensionless variables using the Laplace transform:
\begin{equation*}
  \dz\phi = -\frac{1}{\mu}\sqrt{\frac{\nu}{\tau}}\phi_{zz} \equiv
  \mu\sqrt{\frac{\nu}{\tau}}\grad^2\phi.
\end{equation*}
\end{remark}

We introduced above three classical scaling parameters: the nonlinearity $\eps := \frac{a}{h}$, the dispersion $\mu^2 := \bigl(\frac{h}{\ell}\bigr)^2$ and the inverse Reynolds number also denoted by $\nu$ for the sake of convenience $\nu \rightarrow \frac{\nu}{\sqrt{gh}\;\ell}$. First two parameters are classical in long wave theory \cite{Peregrine1967, bona, Bona1981, Savage, Nwogu1993, BCS, Dutykh2007} while the last one is less classical in this context and it is present due to viscous effects. In the so-called Boussinesq regime it is additionally assumed that nonlinearity and dispersion are of the same order, i.e. the Stokes-Ursell number \cite{Ursell1953} is of the order of unity:
\begin{equation}\label{eq:stokes}
  \mathrm{St} := \frac{\eps}{\mu^2} = \O(1).
\end{equation}
The last assumption $\mathrm{St} \sim 1$ will be systematically used in the following Section. It will allow us to neglect higher order terms since:
\begin{equation*}
  \eps^2 \sim \eps\mu^2 = \O(\mu^4).
\end{equation*}

\section{D2N operator asymptotic expansions}\label{sec:dissB}

In the current Section we simplify the water wave problem \eqref{eq:dynamicsDiss} by expanding asymptotically the D2N map. We apply two different methods which lead to the same result. The first method is called the shape derivative method and consists in expanding the D2N map in the nonlinearity parameter $\eps$. The second method presented in Section \ref{sec:lwave} expands the velocity potential in powers of the dispersion parameter $\mu^2$. It is worth to point out that if the Stokes--Ursell number is of the order of one (the so-called Boussinesq regime \cite{DMII}), i.e. $\varepsilon \sim \mu^2$, both methods lead to the same asymptotic model.

To shorten the notation and make the derivation as clear as possible, we restrict our attention to the 2D case. The horizontal variable is denoted simply by $x$ hereafter. Consequently, we will have to compute an asymptotic approximation to the following expression:
\begin{equation}\label{eq:D2N1d}
  \Dd_{\eps\eta}(\phis) = \left.\bigl(\phi_z - \eps\mu^2\phi_x\eta_x\bigr)\right|_{z=\eps\eta}.
\end{equation}

\subsection{Shape derivative method}\label{sec:shape}

In this section we follow in great lines the study \cite{Lannes2005}. However, the original derivation procedure was greatly simplified to fit our particular purposes.

We know that the velocity potential $\phi(x,z)$ solves the following elliptic boundary-value problem:
\begin{eqnarray}\label{eq:BVPelliptic}
  \phi_{xx} + \frac{1}{\mu^2}\phi_{zz} = 0, && (x,z) \in \R\times[-1, \eps\eta] \\
  \phi = \phis, && z = \eps\eta, \label{eq:freesBC} \\
  \phi_z = \mu\sqrt{\frac{\nu}{\tau}}\phi_{xx}, && z = -1. \label{eq:BVPbot}
\end{eqnarray}
We will find a solution $\phi(x,z;t)$ to this problem in the form:
\begin{equation}\label{eq:epsexp}
  \phi(x,z;t) = \phi^{(0)} + \eps\phi^{(1)} + o(\eps).
\end{equation}
In the following the dependence on time will be omitted to shorten the notation. Now we have to construct zero and first order approximations $\phi^{(0)}$ and $\phi^{(1)}$.

\subsubsection{Solution in the fixed domain}\label{sec:phi0}

It can be easily seen that the function $\phi^{(0)}$ satisfies the following set of equations:
\begin{eqnarray*}
  \phi^{(0)}_{xx} + \frac{1}{\mu^2}\phi^{(0)}_{zz} = 0, && (x,z) \in \Omega_0 = \R\times[-1,0] \\
  \phi^{(0)} = \phis, && z = 0, \\
  \phi^{(0)}_z = \mu\sqrt{\frac{\nu}{\tau}}\phi^{(0)}_{xx}, && z = -1.
\end{eqnarray*}
Since the domain is a strip, we can employ the Fourier transform in the horizontal variable $x$ in order to solve this problem. Throughout this section by $\hat{f}(k)$ we will denote the Fourier transform of a function $f(x)\in L_2(\R)$. In Fourier variable, we then solve the ODE
\begin{equation*}
  \hat{\phi}_{zz}^{(0)} - \mu^2k^2\hat{\phi}^{(0)} = 0,
\end{equation*}
supplemented with boundary conditions $\hat{\phi}^{(0)} = \hat{\varphi}$ at $z = 0$ and $\hat{\phi}^{(0)}_{z} = -\mu k^2 \sqrt{\frac{\nu}{\tau}} \hat{\phi}^{(0)}$ at $z=-1$. After some simple computations we find
\begin{equation*}
\phih^{(0)}(k,z) = \hat{R}_\nu(k,z)[\phish(k)],
\end{equation*}
where the pseudo-differential operator $R_\nu(x,z)[\cdot]$ is defined through its symbol which reads:
\begin{equation*}
  \hat{R}_\nu(k,z)[\cdot] := \frac{\cosh\bigl(\mu|k|(z+1)\bigr)
  -|k|\sqrt{\frac{\nu}{\tau}}\sinh\bigl(\mu|k|(z+1)\bigr)}{\cosh(\mu|k|)
  -|k|\sqrt{\frac{\nu}{\tau}}\sinh(\mu|k|)}\cdot
\end{equation*}

For our purposes it is sufficient to use an approximate operator $\partial_z \hat{R}_\nu(k,z)[\cdot]$ at $z=0$ to the order $\O(\mu^2)$:
\begin{equation*}
\hat{R}_\nu(k,z)[\cdot] = k^2\Bigl(1 - \frac{k^2\mu^2}{3} + \bigl(\mu k^2-\frac{1}{\mu}\bigr)\sqrt{\frac{\nu}{\tau}} + \bigl(\frac{4\mu^2 k^4}{3} - k^2\bigr)\frac{\nu}{\tau} - \mu k^4\bigl(\frac{\nu}{\tau}\bigr)^\frac32 + o(\mu^2)\Bigr),
\end{equation*}
which leads to the following term appearing in the D2N map $\Dd_{\eps\eta}(\phis)$:
\begin{multline*}
  \frac{1}{\mu^2}\phi_z^{(0)}(x,0) = -\phis_{xx} - \frac{\mu^2}{3}\phis_{xxxx} + \frac{1}{\mu}\sqrt{\frac{\nu}{\tau}}(\phis_{xx} + \mu^2\phis_{xxxx}) \\ - \frac{\nu}{\tau}\bigl(\phis_{xxxx} + \frac{\mu^2}{3}\phis_{xxxxxx}\bigr) + \mu\bigl(\frac{\nu}{\tau}\bigr)^\frac32\phis_{xxxxxx} + o(\mu^2).
\end{multline*}
In the last expression we can neglect higher order viscous terms to obtain a simplified representation:
\begin{equation*}
  \frac{1}{\mu^2}\phi_z^{(0)}(x,0) = -\phis_{xx} - \frac{\mu^2}{3}\phis_{xxxx} + \frac{1}{\mu}\sqrt{\frac{\nu}{\tau}}(\phis_{xx} + \mu^2\phis_{xxxx})
  - \frac{\nu}{\tau}\phis_{xxxx} + o(\nu+\mu^2).
\end{equation*}

\subsubsection{The shape derivative}\label{sec:sderiv}

The next step consists in computing the first order approximation $\phi^{(1)}(x,z)$. The main problem is that in a difference with the previous case, the function $\phi^{(1)}$ is defined on a perturbed domain. In order to proceed, we will make a change of variables to work on a fixed domain:
\begin{equation}\label{eq:newvars}
  (x,z)\in \Omega_\eps \longrightarrow (x, y = r(x,z) = \frac{z-\eps\eta}{1+\eps\eta})\in \Omega_0.
\end{equation}
It is easy to see that the Jacobian $J$ of the last transformation is equal to $J = r'_z = \frac{1}{1+\eps\eta}$. Throughout this section all functions defined on the straightened domain $\Omega_0$ will be denoted by tildes, e.g. $\phit(x,y(x,z)) = \phi(x,z)$.

Let $\Phi\in C_0^\infty(\Omega_\eps)$ be a test function. Then, equation \eqref{eq:BVPelliptic} can be rewritten in a weak form:
\begin{equation*}
  \frac{1}{\mu^2}\int\limits_{\Omega_\eps}\phi_z\Phi_z\;dx\,dz +
  \int\limits_{\Omega_\eps}\phi_x\Phi_x\;dx\,dz = 0,
\end{equation*}
or after performing the change of variables \eqref{eq:newvars}:
\begin{equation*}
  \frac{1}{\mu^2}\int\limits_{\Omega_0}\phit_y\psit_y\frac{dx\,dy}{1+\eps\eta} +
  \int\limits_{\Omega_0}\Bigl[(1+\eps\eta)\phit_x - (y+1)\eps\eta_x\phit_y\Bigr]
  \Bigl[\psit_x - \frac{(y+1)\eps\eta_x}{1+\eps\eta}\Bigr]\;dx\,dy = 0.
\end{equation*}
We can substitute into the last integral identity the asymptotic expansion \eqref{eq:epsexp} recasted in new variables:
\begin{equation*}
  \phit = \phit^{(0)} + \eps \phit^{(1)} + o(\eps).
\end{equation*}
At the lowest order it will give us the weak form of the Laplace equation \eqref{eq:BVPelliptic} for $\phit^{(0)}$ which is satisfied identically. At the first order in $\eps$ we obtain the following integral identity:
\begin{multline*}
  \frac{1}{\mu^2}\int\limits_{\Omega_0}\phit^{(1)}_y\psit_y\;dx\,dy +
  \int\limits_{\Omega_0}\phit^{(1)}_x\psit_x\;dx\,dy = \\ = \int\limits_{\Omega_0}\Bigl[\frac{\eta}{\mu^2}\phit^{(0)}_y\psit_y + (y+1)\eta_x(\phit^{(0)}_x\psit_y + \phit^{(0)}_y\psit_x)
  - \eta\phit^{(0)}_x\psit_x \Bigr]\;dx\,dy.
\end{multline*}
Switching back to the differential form and performing some simplifications yield this inhomogeneous elliptic PDE for the first order approximation $\phit^{(1)}$ function:
\begin{equation}\label{eq:pdeEl}
  \frac{1}{\mu^2}\phit^{(1)}_{yy} + \phit^{(1)}_{xx} = \frac{2\eta}{\mu^2}\phit^{(0)}_{yy} + 2(y+1)\eta_x\phit^{(0)}_{xy} + \eta_{xx}(y+1)\phit^{(0)}_y, \quad (x,y)\in \R\times[-1,0].
\end{equation}
This equation should be completed by appropriate boundary conditions which are obtained by expanding corresponding conditions \eqref{eq:freesBC}, \eqref{eq:BVPbot} of the complete problem:
\begin{eqnarray}\label{eq:bc1}
  \phit^{(1)} = 0, && y = 0, \\
  \phit^{(1)}_y - \mu\sqrt{\frac{\nu}{\tau}}\phit^{(1)}_{xx} = \eta\phit^{(0)}_y, && y = -1. \label{eq:bc2}
\end{eqnarray}
In order to solve this problem we will consider the following ansatz for the solution:
\begin{equation}\label{eq:ansatz}
  \phit^{(1)}(x,y) \approx \Phit(x,y) = (y+1)\eta(x)\phit^{(0)}_y(x,y).
\end{equation}
By direct substitution we can check that equation \eqref{eq:pdeEl} and bottom boundary condition \eqref{eq:bc2} are satisfied identically. The free surface condition \eqref{eq:bc1} poses however some problems, since
\begin{equation*}
  \Phit(x,y=0) = \eta(x)\phit^{(0)}_y(x,0) \neq 0.
\end{equation*}
Fortunately, this shortcoming can be easily corrected since the problem \eqref{eq:pdeEl} -- \eqref{eq:bc2} is linear in velocity potential. Consequently, the ansatz \eqref{eq:ansatz} can be corrected to satisfy the free surface condition \eqref{eq:bc1}:
\begin{equation*}
  \phit^{(1)}(x,y) = \Phit(x,y) + \Psit(x,y),
\end{equation*}
where the function $\Psit(x,y)$ is a solution to the following problem:
\begin{eqnarray*}
  \frac{1}{\mu^2}\Psit_{yy} + \Psit_{xx} = 0, && (x,y) \in \R\times[-1,0], \\
  \Psit = -\eta(x)\phit^{(0)}_y, && y = 0, \\
  \Psit_y = 0, && y = -1.
\end{eqnarray*}
The solution to these equations has been constructed in Section~\ref{sec:phi0} (with viscous parameter $\nu = 0$) and an application of operator $R_0$ concludes the construction of the shape derivative:
\begin{equation*}
  \phit^{(1)}(x,y) = (y+1)\eta(x)\phit^{(0)}_y - R_0[\eta(x)\phit^{(0)}_y(x,0))].
\end{equation*}

Now we are able to obtain the asymptotic expansion of the D2N map \eqref{eq:D2N1d}. It can be performed directly in straightened variables $(x,y)$, since one has:
\begin{equation}\label{eq:D2Nscal}
  \Dd_{\eps\eta}(\phis) = \left.\Bigl(\frac{1}{1+\eps\eta}\phit_y - \eps\mu^2\eta_x\phis_x\Bigr)\right|_{y=0} = \left.\Bigl((1-\eps\eta)\phit^{(0)}_y + \eps\phit^{(1)}_y - \eps\mu^2\eta_x\phis_x\Bigr)\right|_{y=0}.
\end{equation}
After substituting expansion \eqref{eq:epsexp} for the velocity potential $\phit$ into the rescaled version of D2N map definition \eqref{eq:D2Nscal} and truncating the expansion at $\O(\eps + \mu^2)$ terms, we obtain the following asymptotic expression which concludes this section:
\begin{multline}\label{eq:D2Nexp}
  \frac{1}{\mu^2}\Dd_{\eps\eta}(\phis) = -\bigl((1+\eps\eta)\phis_x\bigr)_x - \frac{\mu^2}{3}\phis_{xxxx} - \frac{\nu}{\tau}\phis_{xxxx} \\
  + \frac{1}{\mu}\sqrt{\frac{\nu}{\tau}}(\phis_{xx} + \mu^2\phis_{xxxx}) + \O(\mu^3),
\end{multline}
where we retained only the leading order viscous effects $\O(\nu^{\frac12})$ and $\O(\nu)$.

\subsection{Long wave expansion method}\label{sec:lwave}

The developments presented in this section are mainly inspired by the study \cite{Lannes2009}. We recall that we expand here the velocity potential in powers of the dispersion parameter $\mu^2$:
\begin{equation}\label{eq:phiexp}
  \phi(x,z;t) = \sum_{n=0}^{+\infty} \mu^n\phi^{(n)}(x,z;t),
\end{equation}
where the dependence on time can be postponed until considering the dynamic free surface boundary conditions. Substituting expansion \eqref{eq:phiexp} into equations \eqref{eq:BVPelliptic} -- \eqref{eq:BVPbot} leads the following infinite sequence of elliptic problems:
\begin{eqnarray*}
  \phi^{(n)}_{zz} + \phi^{(n-2)}_{xx} = 0, && (x,z) \in \R\times[-1, \eps\eta] \\
  \phi^{(n)} = \delta_{0n}\phis, && z = \eps\eta, \\
  \phi^{(n)}_z = \sqrt{\frac{\nu}{\tau}}\phi^{(n-1)}_{xx}, && z = -1,
\end{eqnarray*}
where $\phi^{(-2)} = \phi^{(-1)} \equiv 0$ by definition and $\delta_{0n}$ is the usual Kronecker symbol (equal to $1$ if $n=0$ and equal to $0$ otherwise). Since we are going to neglect all the terms of the order $\O(\mu^4)$ and higher, we have to compute only first three terms ($\phi^{(0)}$, $\phi^{(1)}$ and $\phi^{(2)}$) in the asymptotic expansion \eqref{eq:phiexp}. After some algebraic computations that we omit here, we obtain the following expressions of the first three solutions in the infinite hierarchy:
\begin{eqnarray}\label{eq:phi0}
  \phi^{(0)}(x,z) &=& \phis(x), \\
  \phi^{(1)}(x,z) &=& \sqrt{\frac{\nu}{\tau}}(z-\eps\eta)\phis_{xx}, \\
  \phi^{(2)}(x,z) &=& -\frac12(z-\eps\eta)(z+\eps\eta+2)\phis_{xx}
  -\frac{\nu}{\tau}\nu(z-\eps\eta)\bigl((1+\eps\eta)\phis_{xx}\bigr)_{xx}. \label{eq:phi2}
\end{eqnarray}
After substituting relations \eqref{eq:phi0} -- \eqref{eq:phi2} into the definition of the D2N map \eqref{eq:D2N1d} and neglecting higher order terms, we derive exactly the same expression \eqref{eq:D2Nexp} for $\Dd_{\eps\eta}(\phis)$ as we obtained in the previous Section~\ref{sec:sderiv}.

Consequently, both derivation methods yield the same result at least for asymptotic orders considered in this study. We have to note that the second method seems to lead to the result in a more direct way. However, the theoretical and practical potential of the shape derivative method may not be fully explored yet. Each method allows to derive models that are relevant in the range of corresponding small parameters: the first one is suitable to derive a family of models with respect to the nonlinearity parameter $\epsilon$, while the second one is relevant for the dispersion parameter $\mu$ at the price to solve a hierarchy of D2N problems.

\section{Dissipative Boussinesq equations}\label{sec:4}

In this section we derive a Boussinesq-type evolution system for viscous long water waves using the asymptotic expansion \eqref{eq:D2Nexp} of the D2N map derived above. Substituting expression \eqref{eq:D2Nexp} into evolution equations \eqref{eq:kinem}, \eqref{eq:dyn} and neglecting higher order terms we obtain the following system of Boussinesq-type equations \cite{DutykhDias2007, Dutykh2008a}:
\begin{equation}\label{eq:massFS}
  \eta_t + \bigl((1+\eps\eta)u\bigr)_x + \frac{\mu^2}{3}u_{xxx}
  + \frac{\nu}{\tau}u_{xxx} =
  2\nu\eta_{xx} + \frac{1}{\mu}\sqrt{\frac{\nu}{\tau}}(u_x+\mu^2 u_{xxx}),
\end{equation}
\begin{equation}\label{eq:momFS}
  u_t + \eps u u_x + \eta_x = 2\nu u_{xx},
\end{equation}
where $u := \phis_x$ is the velocity defined at the free surface.

We can further improve dispersive properties of the derived Boussinesq system \eqref{eq:massFS}, \eqref{eq:momFS} if we change the velocity variable to the depth averaged velocity \cite{Peregrine1967}, for example. The choice of the velocity variable is far from being unique \cite{BS, Nwogu1993, BCS}, but we prefer not to enter into the debate.

The asymptotic representation \eqref{eq:phi0} -- \eqref{eq:phi2} can be averaged over the water depth to provide us the corresponding expression for the depth-averaged velocity potential:
\begin{equation*}
  \phib = \phis - \mu\sqrt{\frac{\nu}{\tau}}\phis_{xx} + \frac{\mu^2}{3}\phis_{xx} + \O(\mu^3 + \nu\mu^2).
\end{equation*}
In order to invert this relation and, thus, express the velocity potential at the free surface $\phis$ in terms of $\phib$, it is better to work with Fourier transforms \cite{BCS}:
\begin{equation*}
  \phibh = \Bigl(1 + \mu\sqrt{\frac{\nu}{\tau}}k^2 - \frac{\mu^2}{3}k^2\Bigr)\phish + \O(\mu^3).
\end{equation*}
The last relation can be easily inverted within the same asymptotic accuracy:
\begin{equation*}
  \phish = \Bigl(1 + \mu\sqrt{\frac{\nu}{\tau}}k^2 - \frac{\mu^2}{3}k^2\Bigr)^{-1}\phibh + \O(k^4) = \Bigl(1 - \mu\sqrt{\frac{\nu}{\tau}}k^2 + \frac{\mu^2}{3}k^2\Bigr)\phibh + \O(k^4).
\end{equation*}
In physical space we obtain the requested relation between $\phis$ and $\phib$:
\begin{equation*}
  \phis = \phib + \mu\sqrt{\frac{\nu}{\tau}}\phib_{xx} - \frac{\mu^2}{3}\phib_{xx} + \O(\mu^3).
\end{equation*}
The direct substitution of the last relation into governing equations \eqref{eq:momFS}, \eqref{eq:massFS} and inverting the Laplace equation leads the following set of Boussinesq type equations which represent a viscid generalization of the celebrated Peregrine system over the flat bottom \cite{Peregrine1967}:
\begin{equation}\label{eq:vper1}
  \eta_t + \bigl((1+\eps\eta)\ub\bigr)_x = 2\nu\eta_{xx} +
  \frac{1}{\mu}\sqrt{\frac{\nu}{\pi}}\int\limits_0^t\frac{(\ub_x -
  \frac{\mu^2}{3}\ub_{xxx})(x,s)}{\sqrt{t-s}}\,ds,
\end{equation}
\begin{equation}\label{eq:vper2}
  \ub_t + \eps\ub\ub_x + \eta_x - \frac{\mu^2}{3}\ub_{xxt} = 2\nu\ub_{xx}
  -\pd{}{t}\Bigl(\frac{1}{\mu}\sqrt{\frac{\nu}{\pi}}\int\limits_0^t
  \frac{\mu^2\ub_{xx}(x,s)}{\sqrt{t-s}}\,ds\Bigr),
\end{equation}
where $\ub := \phib_x$ is the depth-averaged velocity variable. To our knowledge this is the first dissipative Boussinesq system which contains a half-order derivative and a half-order integral in the same time.

\subsection{Dispersion relation analysis}

Let us consider the linearized version of the viscous Peregrine system \eqref{eq:vper1}, \eqref{eq:vper2} with dimensionless parameters $\eps$, $\mu^2$ set for convenience to the unity (it is always possible by choosing the appropriate scaling):
\begin{equation}\label{eq:lp1}
  \eta_t + \ub_x = 2\nu\eta_{xx} +
  \sqrt{\frac{\nu}{\pi}}\int\limits_0^t\frac{(\ub_x -
  \frac{1}{3}\ub_{xxx})(x,s)}{\sqrt{t-s}}\,ds,
\end{equation}
\begin{equation}\label{eq:lp2}
  \ub_t + \eta_x - \frac{1}{3}\ub_{xxt} = 2\nu\ub_{xx}
  -\pd{}{t}\Bigl(\sqrt{\frac{\nu}{\pi}}\int\limits_0^t
  \frac{\ub_{xx}(x,s)}{\sqrt{t-s}}\,ds\Bigr).
\end{equation}
Equations \eqref{eq:lp1}, \eqref{eq:lp2} can be analyzed if we apply the Fourier transform in space $x\to k$ and the Laplace transform \cite{Widder1946} in time $t\to \tau$ reducing these PDEs to a system of two linear equations. This analysis is relatively classical, consequently, we do not enter into algebraic details. The solvability condition for this system gives the required dispersion relation between the frequency $\tau$ and the wavenumber $k$:
\begin{equation}\label{eq:dispRelb}
  (\tau + 2\nu k^2)\Bigl(\tau\bigl(1+\frac{k^2}{3}\bigr)+2\nu k^2 -   
  \sqrt{\tau\nu}k^2\Bigr) + k^2\Bigl(1-\sqrt{\frac{\nu}{\tau}}   
  \bigl(1+\frac{k^2}{3}\bigr)\Bigr) = 0.
\end{equation}
The last relation is a polynomial equation of the degree five if we make a change of variables $\tau = z^2$. There is no general closed-form solution to this equation. However, it can be efficiently solved numerically for any given combination of parameters.

For comparison, the dispersion relation of the free surface Euler equations (the fluid is supposed to be perfect) takes the following classical form \cite{Mei1994, Whitham1999}:
\begin{equation}\label{eq:EulerDisp}
  \tau^2 + k\tanh k = 0.
\end{equation}
Below we make a comparison between the last relation with the dispersion relation \eqref{eq:dispRelb} to viscous Peregrine equations \eqref{eq:vper1}, \eqref{eq:vper2}.

\subsubsection{Numerical results and discussion}

In this section we assess the performance of the derived above viscous Peregrine system in approximating the dispersive properties of the full Euler equations for a perfect fluid with a free surface. The dispersion relation analysis allows also to study the dissipative properties of the proposed model. For this purpose we consider another quantity of interest which is called the phase velocity defined as:
\begin{equation*}
  c(k) := \frac{\tau(k)}{k} = c_r(k) + ic_i(k),
\end{equation*}
where the imaginary part $c_i(k)$ is responsible for the wave propagation phenomena and the real part $c_r(k)$ describes the dissipative properties of the system (small perturbations are damped if $c_r<0$ and amplified otherwise).

The wave frequency $\tau$ is found for each value of the wavenumber $k$ by solving numerically (Newton-type iterations initialized with relation \eqref{eq:EulerDisp}) equation \eqref{eq:dispRelb}. In this way we obtained numerical results presented below. For the full Euler equations we use the exact analytical relation \eqref{eq:EulerDisp}. Finally, the phase speed is obtained after a simple division operation by the wavenumber $k$.

\begin{remark}\label{rem:visc}
The choice of the eddy viscosity for practical simulations of water waves is not obvious. However, in various experimental and theoretical studies researchers independently come to the conclusion that a value of $\nu \approx 10^{-3}$ $m^2/s$ fits very well available data \cite{Tuck, Bona1981, Tuck2007}. The most recent experimental study of plunging breakers confirms this value again \cite{Tian2010}. Consequently, we retain this value for our numerical illustrations as well. For comparison, the molecular viscosity of the water is of the order $\nu \approx 10^{-6}$ $m^2/s$ which is too small to model the energy dissipation phenomena in a laboratory wave tank, for example. Hence, this molecular viscosity should be replaced by its effective value.
\end{remark}

\begin{figure}
  \centering
  \subfigure[\em\small Zoom on long waves]%
  {\includegraphics[width=0.49\textwidth]{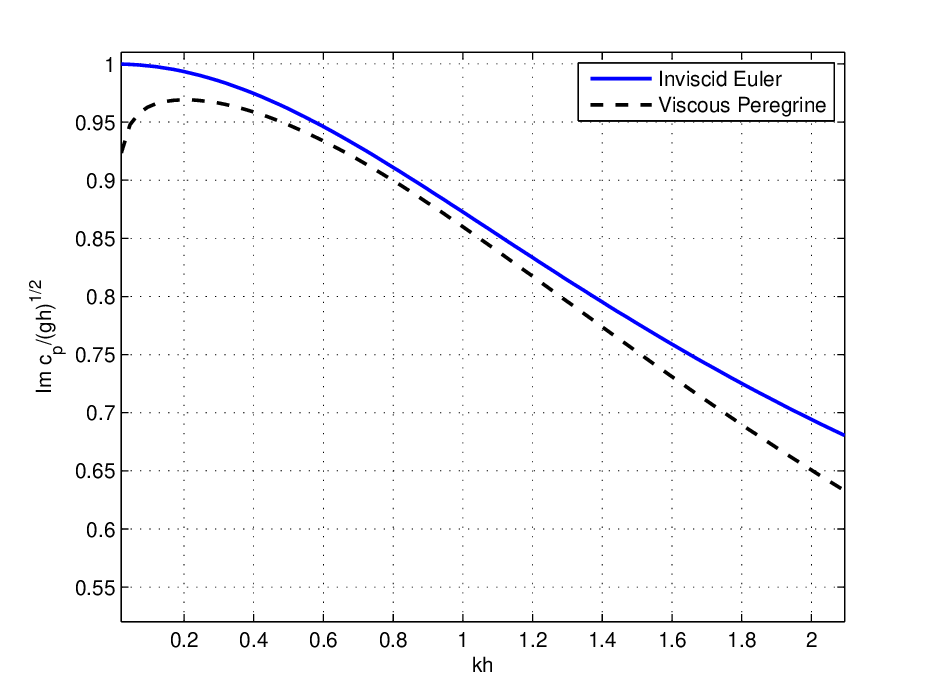}}
  \subfigure[\em\small Long and short waves]%
  {\includegraphics[width=0.49\textwidth]{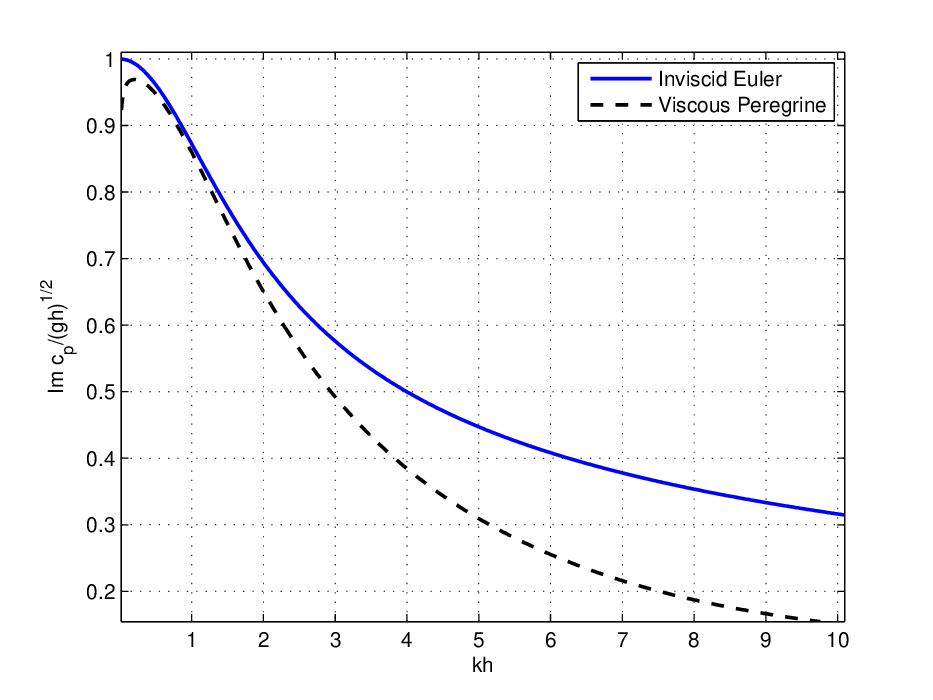}}
  \caption{\em\small Imaginary part of the phase velocity with zoom on long waves region on the left image. The blue solid line corresponds to the dispersion relation of the full Euler equations while the black dashed line represents viscous Peregrine system.}
  \label{fig:imag}
\end{figure}

\begin{figure}
  \centering
  \subfigure[\em\small Zoom on long waves]%
  {\includegraphics[width=0.49\textwidth]{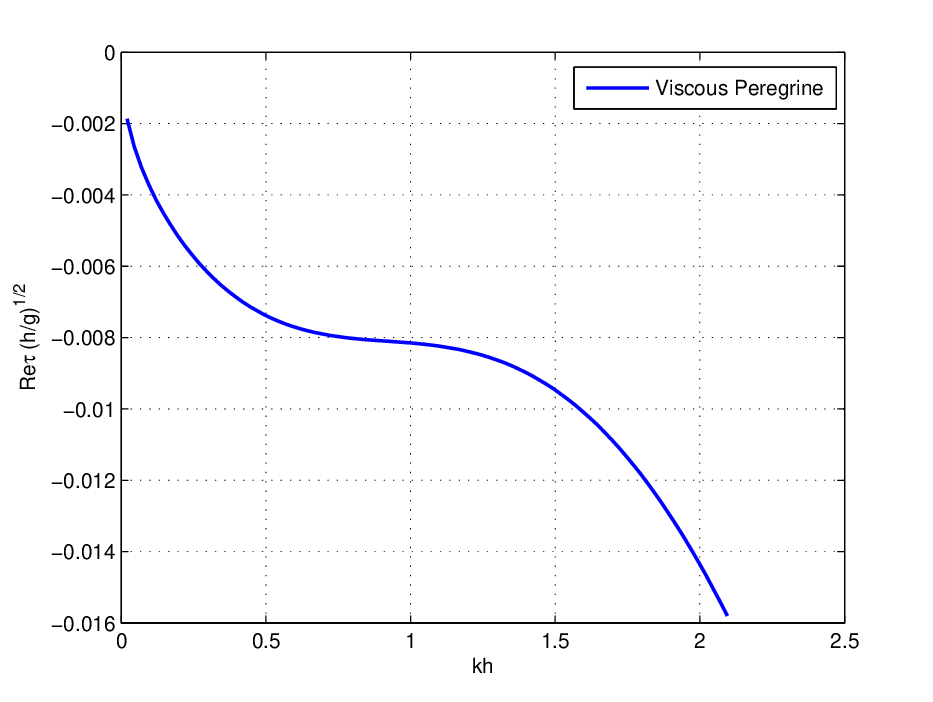}}
  \subfigure[\em\small Long and short waves]%
  {\includegraphics[width=0.49\textwidth]{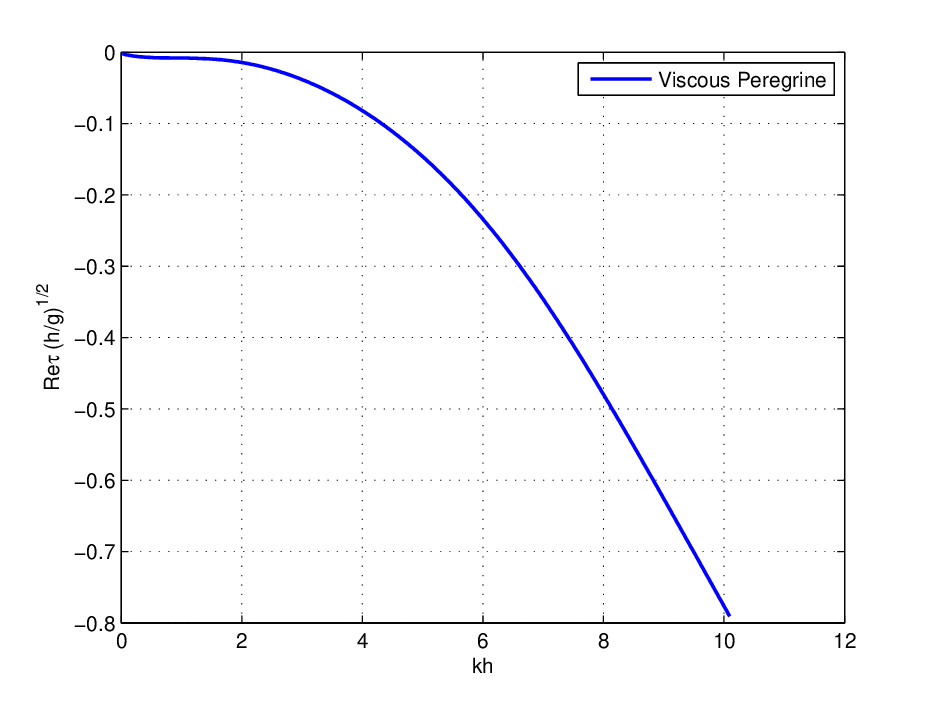}}
  \caption{\em\small Real part of the wave frequency $\tau$ with zoom on long waves region on the left image. The real part of the phase velocity of full Euler equations is identically equal to zero since the fluid is assumed to be perfect.}
  \label{fig:real}
\end{figure}

On Figure~\ref{fig:imag} we show the imaginary part of the phase velocity for Euler and viscous Peregrine equations. For our numerical computations we fixed the viscosity parameter $\nu$ to be equal to $10^{-3}$ (see Remark~\ref{rem:visc}). A zoom on long waves region is shown on Figure~\ref{fig:imag}(a) where we can see that the wave propagation is slightly slown down by viscous effects. Mathematically this effect is ascribed to nonlocal terms which are more important in magnitude for small wavenumbers $kh$. We can conclude that for $kh < 2$ the viscous Peregrine system provides a reasonnable approximation to full Euler equations. The behaviour of the phase velocity for shorter waves is shown on Figure~\ref{fig:imag}(b). The real part of the wave frequency $\tau$ is depicted on Figure~\ref{fig:real}. On the left image in Figure~\ref{fig:real}(a) one can see a zoom on the long wave region. The apparent transient behaviour at $kh\approx 1$ corresponds to the change between the region where the nonlocal terms are dominant ($kh < 1$ for the present choice of $\nu$) to the short wave region dominated by local diffusion ($\tau \approx -2\nu k^2$ for $kh>1$).

\section{Conclusions}\label{sec:concl}

In the present study we investigate further the derivation methods of dissipative Boussinesq equations \cite{Dutykh2007, Dutykh2008a}. First, we recast the visco-potential formulation \cite{Liu2004, DutykhDias2007} using the normal velocity at the free surface which is given by the so-called Dirichlet-to-Neumann map in the spirit of the Petrov-Zakharov formulation \cite{Petrov1964, Zakharov1968}. Then, two completely different approaches have been employed to simplify this complete visco-potential formulation. The first method is called the shape derivative method \cite{Lannes2005} which expands the D2N map in powers of the nonlinearity parameter $\eps$. The second method is more classical \cite{Madsen1991, Madsen1998, Lannes2009} and consists in expanding the velocity potential $\phi$ and, consequently, the D2N map in powers of the dispersion (or shallowness) parameter $\mu^2$. Both methods lead to the same asymptotic expression for the D2N map and thus, to the same dissipative Boussinesq type system. The shape derivative method seems to involve more computational work. However, we prefer to keep both methods in our exposition. The main reason is that in the absence of the Stokes number assumption ($\eps \sim \mu^2$), two methods are fundamentally different since they involve expansions in different small parameters (nonlinearity and dispersion). Consequently, the shape derivative method is fully dispersive, while the long wave expansion is fully nonlinear if no additional assumptions are made on the magnitude of the second parameter (dispersion and nonlinearity, respectively). Moreover, we came up with two new Boussinesq type systems, one of them being a viscid generalization of classical Peregrine system \cite{Peregrine1967}. Finally, the dispersion relation properties of 
the derived system were discussed as well.

In future investigations the decay rate of solutions to new systems at infinite times $t\to+\infty$ will be studied. These asymptotic estimations can be supported by several numerical simulations of a solitary wave decay rate. The effect of new viscous terms is to be revealed in future studies.

\section*{Acknowledgement}

D.~Dutykh acknowledges the support from French Agence Nationale de la Recherche, project MathOcean (Grant ANR-08-BLAN-0301-01) along with the support from ERC under the research project ERC-2011-AdG 290562-MULTIWAVE. Also D.~Dutykh would like to acknowledge the support from the University of Picardie Jules Verne during his visity to Amiens. The authors would like to thank Herv\'e Le Meur and Professors Min Chen and Fr\'ed\'eric Dias for helpful discussions on viscous Boussinesq equations.

\bibliography{biblio}
\bibliographystyle{abbrv}

\end{document}